\documentclass[pre,a4paper,oneside,twocolumn]{revtex4}

\usepackage{color} 
\usepackage[normalem]{ulem} % Pacote para comando de riscar texto: \sout{texto}

\usepackage{graphicx}

\begin{document}

\title{Zipf's law and urban scaling: Hypotheses towards a Unified Urban Theory}

\author{Fabiano L.\ Ribeiro}
\affiliation{Department of Physics (DFI), Federal University of Lavras (UFLA), Lavras MG, Brazil}
\author{Jose Lobo}
\affiliation{School of Sustainability, Arizona State University, 800 South Cady Mall, Tempe, AZ 85281, USA}
\author{Diego Rybski}
\email{ca-dr@rybski.de}
\affiliation{Potsdam Institute for Climate Impact Research -- PIK, Member of Leibniz Association, P.O.\ Box 601203, 14412 Potsdam, Germany}
\affiliation{Department of Environmental Science Policy and Management, University of California Berkeley, 130 Mulford Hall \#3114, Berkeley, CA 94720, USA}
\affiliation{Complexity Science Hub Vienna, Josefst\"adterstrasse 39, A-1090 Vienna, Austria}

\date{\today}

\begin{abstract}
We propose hypotheses describing the empirical finding of an association between the exponents of urban GDP scaling and Zipf's law for cities.
These hypotheses represent various combinations of directional or reciprocal causal links between the two phenomena and include inter- and intra-city processes. 
Future theories and models can be motivated with and categorized according to these hypotheses.
This paper intends to stimulate the discussion around the processes behind these phenomena and pave the way to a Unified Urban Theory.
\end{abstract}

\maketitle

In 1975 Sveikauskas formulated 
\textit{``($\cdots$) the observed relationship between city size and productivity could come about either because city size itself causes productivity to be high or because individual cities systematically grow to large size because they are already more productive''} \cite{SveikauskasL1975}.
He is referring to what nowadays is researched as \emph{urban scaling}, namely the phenomenon that bigger cities are proportionally more productive and wealthier. 
It is characterized by a larger than one scaling exponent that relates socio-economic metrics and population size \cite{BettencourtLHKW2007}.

\begin{figure}
\centering
\includegraphics[width=\columnwidth]{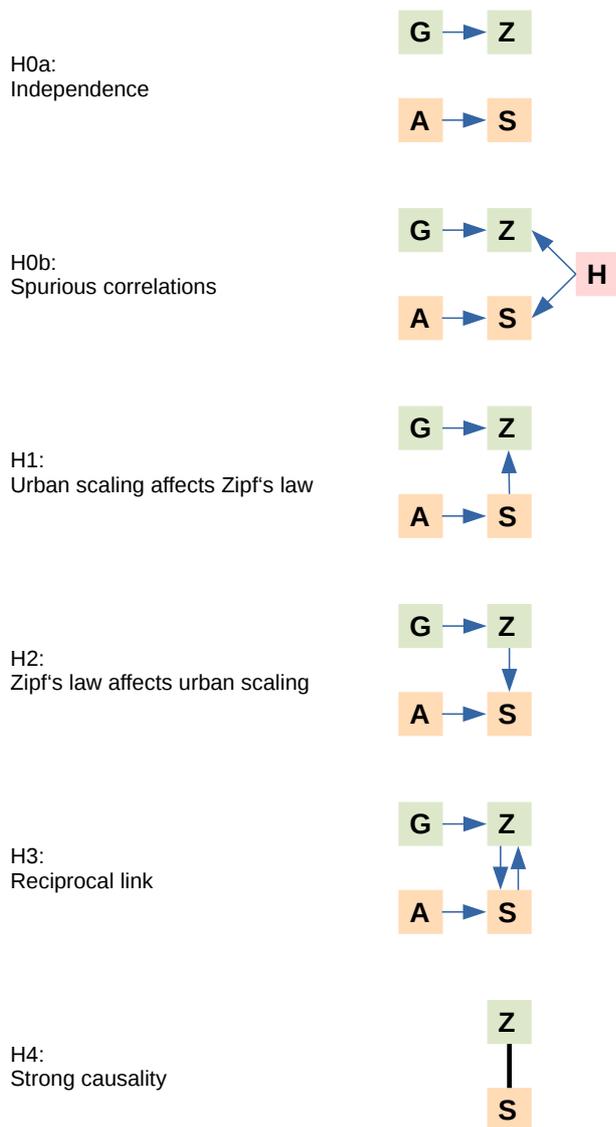}
\caption{Illustration of the hypotheses. The letters have the following meanings.
`G' for Gibrat's law, `Z' for Zipf's law, `A' for agglomeration effects, `S' for urban scaling, and `H' for hidden variable.
Please note that Gibrat's law and agglomeration effects are mentioned as examples and can be replaced by other processes.
}
\label{fig:illu}
\end{figure}

Recently, an association between urban GDP scaling and Zipf's law was reported \cite{RibeiroOMMKR2021}, a finding that could represent a step towards solving Sveikauskas' conundrum.
\emph{Zipf's law for cities} \cite{AuerbachF1913,ZipfGK2012,BerryOK2012,GabaixX1999} states that city size is proportional to the inverse of the rank, which corresponds to a power-law probability density function with an exponent close to 2.
Urban scaling and Zipf's law are independently well researched but opinions diverge on the question whether they are related or not.
Ribeiro et al. (2021) \cite{RibeiroOMMKR2021} also provide a theoretical derivation but make no statement about any causal link.
In the following we want to propose and put up for discussion hypotheses (Fig.~\ref{fig:illu}) on the presence or absence of causality between the two phenomena. 
Perspectively, such a discussion could pave the way to a \emph{Unified Urban Theory} (UUT).

\textbf{Hypothesis H0a: Independence.}
This hypothesis states that both laws are independent. 
On the one hand, Zipf's law is e.g.\ due to cities growing independently and growth rates are independent of the sizes, an assumption also known \textit{Gibrat's law} \cite{SuttonJ1997}.
On the other hand, urban scaling can e.g.\ be explained by increasing human interactions and agglomeration effects within cities with increasing size \cite{BettencourtLMA2013}.
This hypothesis, that is also based on the premise that both laws are determined by intra-city mechanisms, is not in agreement with the recent findings and only included as rival hypothesis.

\textbf{Hypothesis H0b: Spurious correlations.}
This hypothesis states that another factor influences both laws and without knowledge about this factor, Zipf's law and urban scaling seem related but there is no direct causal link between them.
Ribeiro et al. \cite{RibeiroOMMKR2021} already mention a potential factor, namely economic development.
Countries with higher GDP tend to exhibit Zipf exponents closer to the ideal value and urban GDP scaling closer to linearity.
It is also included as rival hypothesis.

\textbf{Hypothesis H1: Urban scaling affects Zipf's law.}
This hypothesis states that urban scaling and Zipf's law emerge due to processes mentioned in H0a -- in addition there is an influence from urban scaling on Zipf's law.
Obviously, an alteration of the city size distribution requires migration (an inter-city mechanism).
A possible reason for people to move from one city to another is the aspiration for a better livelihood in large cities.
Population of smaller cities is attracted by more wealth and culture in larger cities (generated/enhanced by agglomeration effects), which leads to an adjustment of the Zipf distribution.

\textbf{Hypothesis H2: Zipf's law affects urban scaling.}
This hypothesis states that Zipf's law and urban scaling emerge due to processes mentioned in H0a -- in addition there is an influence from Zipf's law on urban scaling (complementary to H1).
A possible mechanism could be the following \cite{RibeiroOMMKR2021}.
Particularly steep city size distributions come along with extreme metropolises.
As these very large cities concentrate a lot of population, they also agglomerate diverse economic sectors and businesses.
Accordingly larger is the urban scaling exponent.
In contrast, more balanced city size distributions include several large cities and the diverse economic activities are distributed among them, which implies an urban scaling exponent closer to linearity.

\textbf{Hypothesis H3: Reciprocal link.}
This hypothesis states that Zipf's law and urban scaling emerge due to processes mentioned in H0a.
Combining H1 and H2, Zipf's law influences urban scaling and vice versa.
A back-propagation process adjusts urban scaling and Zipf's law.

\textbf{Hypothesis H4: Strong causality.}
This hypothesis states that one phenomenon actually causes the other.
According to the hypotheses H1-H3 Zipf's law and urban scaling emerge due to other processes and possible causal interactions among the two lead to an adjustment of their exponents.
H4 goes beyond this, meaning that one or both phenomena would not exist without the respective other one.
Similar to the hypotheses H1-H3, this causality can go in either or both directions.

% Equilibrium
The hypotheses H1-H3 require an undefined counteracting process in order to keep the system in an equilibrium.
Here we do not want to speculate about the nature of this counteracting process -- it is a matter of future research.
Nevertheless, we can discuss under which conditions adjustments of Zipf's law or urban scaling can take place.
First, in countries of ongoing urbanization the urban system is not mature yet and the two laws need some time to reach an equilibrium.
Second, in countries with a mature urban system, perturbation (e.g.\ housing crisis) can affect one or both laws so that the respective other needs to adjust.

% Testing
Since it is practically impossible to perform experiments with real-world urban systems, testing of the hypotheses will be challenging.
However, the above mentioned urbanization trends and perturbation might represent a statistical test-bed.
In addition, modeling of candidate processes can confirm or rule-out their capability of contributing to any of the hypotheses.

% Relevance
Which of the above hypotheses is actually true is not only of scientific relevance but also important for real-world urban and regional planning.
A deeper understanding could enable to develop policies to actually control Zipf's law and urban scaling.
In other words, identifying optimal city size distributions could support to create public policies to improve countries development.

Zipf's law for cities and urban scaling are two out of seven fundamental regularities listed in Michael Batty's book \emph{The New Science of Cities} \cite{BattyM2013NewScience}. 
He also speculates that two or more of those seven laws might be expressions of the same.
Ribeiro et al.\ (2021) \cite{RibeiroOMMKR2021} find correlations between the exponents of two of them, which calls to ask for a possible causal connection.
We hope that the above formulated hypotheses stimulate the scientific discussion around causal interactions among those laws and possibly lead to a UUT.

To deal with all the problems that come with urban intensification, such as high density, traffic, infrastructure saturation, it is a mandate to develop a quantitative, systematic, and unified theory.
Such a theory does not explain either -- urban scaling or Zipf's law -- but covers both and possibly others, e.g.\ mentioned by Batty \cite{BattyM2013NewScience}.
As illustrated by Zipf's law, such a theory not only includes processes within a city but also among cities (intra-city and inter-city processes, respectively).
UUT could represent an important constituent of a new \emph{urban science} \cite{Loboetal2020}.
Identifying general laws and developing a unified theory is essential because it will allow policy-makers to take decisions and manage the cities more efficiently.

%\newpage
%\clearpage

\begin{thebibliography}{11}
\expandafter\ifx\csname natexlab\endcsname\relax\def\natexlab#1{#1}\fi
\expandafter\ifx\csname bibnamefont\endcsname\relax
  \def\bibnamefont#1{#1}\fi
\expandafter\ifx\csname bibfnamefont\endcsname\relax
  \def\bibfnamefont#1{#1}\fi
\expandafter\ifx\csname citenamefont\endcsname\relax
  \def\citenamefont#1{#1}\fi
\expandafter\ifx\csname url\endcsname\relax
  \def\url#1{\texttt{#1}}\fi
\expandafter\ifx\csname urlprefix\endcsname\relax\def\urlprefix{URL }\fi
\providecommand{\bibinfo}[2]{#2}
\providecommand{\eprint}[2][]{\url{#2}}

\bibitem[{\citenamefont{Sveikauskas}(1975)}]{SveikauskasL1975}
\bibinfo{author}{\bibfnamefont{L.}~\bibnamefont{Sveikauskas}},
  \bibinfo{journal}{Q. J. Econ.} \textbf{\bibinfo{volume}{89}},
  \bibinfo{pages}{393} (\bibinfo{year}{1975}).

\bibitem[{\citenamefont{Bettencourt et~al.}(2007)\citenamefont{Bettencourt,
  Lobo, Helbing, K{\"u}hnert, and West}}]{BettencourtLHKW2007}
\bibinfo{author}{\bibfnamefont{L.~M.~A.} \bibnamefont{Bettencourt}},
  \bibinfo{author}{\bibfnamefont{J.}~\bibnamefont{Lobo}},
  \bibinfo{author}{\bibfnamefont{D.}~\bibnamefont{Helbing}},
  \bibinfo{author}{\bibfnamefont{C.}~\bibnamefont{K{\"u}hnert}},
  \bibnamefont{and} \bibinfo{author}{\bibfnamefont{G.~B.} \bibnamefont{West}},
  \bibinfo{journal}{Proc. Natl. Acad. Sci. U. S. A.}
  \textbf{\bibinfo{volume}{104}}, \bibinfo{pages}{7301} (\bibinfo{year}{2007}).

\bibitem[{\citenamefont{Ribeiro et~al.}(2021)\citenamefont{Ribeiro, Oehlers,
  Moreno-Monroy, Kropp, and Rybski}}]{RibeiroOMMKR2021}
\bibinfo{author}{\bibfnamefont{H.~V.} \bibnamefont{Ribeiro}},
  \bibinfo{author}{\bibfnamefont{M.}~\bibnamefont{Oehlers}},
  \bibinfo{author}{\bibfnamefont{A.~I.} \bibnamefont{Moreno-Monroy}},
  \bibinfo{author}{\bibfnamefont{J.~P.} \bibnamefont{Kropp}}, \bibnamefont{and}
  \bibinfo{author}{\bibfnamefont{D.}~\bibnamefont{Rybski}},
  \bibinfo{journal}{PLoS One} \textbf{\bibinfo{volume}{16}},
  \bibinfo{pages}{e0245771} (\bibinfo{year}{2021}).

\bibitem[{\citenamefont{Auerbach}(1913)}]{AuerbachF1913}
\bibinfo{author}{\bibfnamefont{F.}~\bibnamefont{Auerbach}},
  \bibinfo{journal}{Petermanns Geogr.\ Mitteilungen}
  \textbf{\bibinfo{volume}{59}}, \bibinfo{pages}{73} (\bibinfo{year}{1913}).

\bibitem[{\citenamefont{Zipf}(2012)}]{ZipfGK2012}
\bibinfo{author}{\bibfnamefont{G.~K.} \bibnamefont{Zipf}},
  \emph{\bibinfo{title}{Human Behavior and the Principle of Least Effort: An
  Introduction to Human Ecology}} (\bibinfo{publisher}{Martino Publishing},
  \bibinfo{address}{Manfield Centre, CT}, \bibinfo{year}{2012}), ISBN
  \bibinfo{isbn}{978-1614273127}, \bibinfo{note}{(Reprint of 1949 Edition)}.

\bibitem[{\citenamefont{Berry and Okulicz-Kozaryn}(2012)}]{BerryOK2012}
\bibinfo{author}{\bibfnamefont{B.~J.~L.} \bibnamefont{Berry}} \bibnamefont{and}
  \bibinfo{author}{\bibfnamefont{A.}~\bibnamefont{Okulicz-Kozaryn}},
  \bibinfo{journal}{Cities} \textbf{\bibinfo{volume}{29}}, \bibinfo{pages}{S17}
  (\bibinfo{year}{2012}).

\bibitem[{\citenamefont{Gabaix}(1999)}]{GabaixX1999}
\bibinfo{author}{\bibfnamefont{X.}~\bibnamefont{Gabaix}}, \bibinfo{journal}{Am.
  Econ. Rev.} \textbf{\bibinfo{volume}{89}}, \bibinfo{pages}{129}
  (\bibinfo{year}{1999}).

\bibitem[{\citenamefont{Sutton}(1997)}]{SuttonJ1997}
\bibinfo{author}{\bibfnamefont{J.}~\bibnamefont{Sutton}}, \bibinfo{journal}{J.
  Econ. Lit.} \textbf{\bibinfo{volume}{35}}, \bibinfo{pages}{40}
  (\bibinfo{year}{1997}), \urlprefix\url{https://www.jstor.org/stable/2729692}.

\bibitem[{\citenamefont{Bettencourt}(2013)}]{BettencourtLMA2013}
\bibinfo{author}{\bibfnamefont{L.~M.~A.} \bibnamefont{Bettencourt}},
  \bibinfo{journal}{Science} \textbf{\bibinfo{volume}{340}},
  \bibinfo{pages}{1438} (\bibinfo{year}{2013}).

\bibitem[{\citenamefont{Batty}(2013)}]{BattyM2013NewScience}
\bibinfo{author}{\bibfnamefont{M.}~\bibnamefont{Batty}},
  \emph{\bibinfo{title}{The New Science of Cities}} (\bibinfo{publisher}{{MIT}
  Press}, \bibinfo{address}{Cambridge, MA}, \bibinfo{year}{2013}), ISBN
  \bibinfo{isbn}{978-0262019521}.

\bibitem[{\citenamefont{Lobo et~al.}(2020)\citenamefont{Lobo, Alberti,
  Allen-Dumas, Arcaute, Barthelemy, {Bojórquez Tapia}, Brail, Bettencourt,
  Beukes, Chen et~al.}}]{Loboetal2020}
\bibinfo{author}{\bibfnamefont{J.}~\bibnamefont{Lobo}},
  \bibinfo{author}{\bibfnamefont{M.}~\bibnamefont{Alberti}},
  \bibinfo{author}{\bibfnamefont{M.}~\bibnamefont{Allen-Dumas}},
  \bibinfo{author}{\bibfnamefont{E.}~\bibnamefont{Arcaute}},
  \bibinfo{author}{\bibfnamefont{M.}~\bibnamefont{Barthelemy}},
  \bibinfo{author}{\bibfnamefont{L.~A.} \bibnamefont{{Bojórquez Tapia}}},
  \bibinfo{author}{\bibfnamefont{S.}~\bibnamefont{Brail}},
  \bibinfo{author}{\bibfnamefont{L.~M.~A.} \bibnamefont{Bettencourt}},
  \bibinfo{author}{\bibfnamefont{A.}~\bibnamefont{Beukes}},
  \bibinfo{author}{\bibfnamefont{W.-Q.} \bibnamefont{Chen}},
  \bibnamefont{et~al.}, \bibinfo{journal}{Report submitted to the NSF on the
  Present State and Future of Urban Science}  (\bibinfo{year}{2020}).

\end{thebibliography}
\end{document}